\documentclass[preprint2]{aastex62}
\usepackage{graphicx}
\usepackage{rotating}
\usepackage{amssymb}
\usepackage{multirow}
\newcommand{\mlc}{\multicolumn}
\newcommand{\mlr}{\multirow}

\received{December 16, 2019}
\accepted{January 19, 2020}

\shorttitle{High Resolution Spectral Indices}
\shortauthors{Rodriguez-Merino et al.}

\begin{document}

\title{High Resolution Spectral Line Indices Useful for the Analysis of Stellar Populations}

\correspondingauthor{Lino H. Rodr{\'\i}guez-Merino}
\email{lino@inaoep.mx}

\author{Lino H. Rodr{\'\i}guez-Merino}
\affil{Instituto Nacional de Astrof{\'\i}sica, \'Optica y Electr\'onica \\
Luis Enrique Erro 1, Tonantzintla, Puebla, C.P. 72840, Mexico}

\author{Y. D. Mayya}
\affiliation{Instituto Nacional de Astrof{\'\i}sica, \'Optica y Electr\'onica \\
Luis Enrique Erro 1, Tonantzintla, Puebla, C.P. 72840, Mexico}

\author{Paula R. T. Coelho}
\affiliation{Universidad de S\~ao Paulo, Instituto de Astronomia, Geof\'isica e Ci\^encias Atmosf\'ericas \\
Rua do Mat\~ao 1226, 05508-090, S\~ao Paulo, Brazil}

\author{Gustavo Bruzual}
\affiliation{Instituto de Radioastronom{\'\i}a y Astrof{\'\i}sica, Universidad Nacional Aut\'onoma de M\'exico, \\
Morelia, Michoac\'an, C.P. 58090, Mexico}

\author{St\'ephane Charlot}
\affiliation{Sorbonne Universit\'e, CNRS, UMR7095, Institut d'Astrophysique de Paris, F-75014, Paris, France}

\author{Esperanza Carrasco}
\affil{Instituto Nacional de Astrof{\'\i}sica, \'Optica y Electr\'onica \\
Luis Enrique Erro 1, Tonantzintla, Puebla, C.P. 72840, Mexico}

\author{Armando Gil de Paz}
\affiliation{Universidad Complutense de Madrid, Departamento de F{\'\i}sica de la Tierra y Astrof{\'\i}sica, \\Instituto de F\'isica de Part\'iculas y del Cosmos IPARCOS, E-28040, Madrid, Spain}

\begin{abstract}
The well-known age-metallicity-attenuation degeneracy does not permit unique and good estimates of basic 
parameters of stars and stellar populations. The effects of dust can be avoided using spectral line indices, 
but current methods have not been able to break the age-metallicity degeneracy. Here we show that using at 
least two new spectral line indices defined and measured on high-resolution ({\it R}= 6000) spectra of a 
signal-to-noise ratio (S/N) $\ge 10$ one gets unambiguous estimates of the age and metallicity of 
intermediate to old stellar populations. Spectroscopic data retrieved with new astronomical facilities, 
e.g., X-shooter, MEGARA, and MOSAIC, can be employed to infer the physical parameters of the emitting source 
by means of spectral line index and index--index diagram analysis.
\end{abstract}


\section{Introduction} 
\label{sec:intro}
New astronomical facilities allow us to obtain stellar spectra at very high resolution. Therefore, it 
becomes necessary to develop new methods or update existing tools used to study low-resolution spectra 
and render them useful for analyzing the data provided by the new instruments. One of the tools most 
frequently used in the analysis of spectroscopic data is the study of spectral line indices. 

The use of spectral lines to understand the physical properties of stars and stellar populations started 
in the late 1960s of the past century \citep[see][]{Spinrad69,Mould78,Faber85,Buzzoni92,Chavez96}. The 
work carried out during two decades by astronomers at Lick Observatory led to the definition of a set of 
21 spectral line indices \citep[see][]{Worthey94,Worthey97,Trager98} which has been widely used to study 
stars, star clusters, and galaxies \citep{Parikh19,Sharina19}. The Lick set of indices was defined using 
spectra of the highest resolution available at the time ($\sim 8$\,{\AA}), each index being $\sim 30$\,{\AA} 
wide, making them immune to dust effects. However, the age-metallicity degeneracy cannot be easily avoided 
\citep{Rose85, Proctor02,Kaviraj07}. In general, inside the 30\,{\AA} width defining each index there are 
several spectral features that behave in different ways as a function of time and metallicity, preventing 
the Lick indices from breaking this degeneracy \citep{Worthey94b}. In fact, the strength of each spectral 
line is under the effect of the age-metallicity degeneracy. Index--index diagrams have helped to 
overcome degeneracy problems; for example, these diagrams separated effects due to the overabundance 
of $\alpha$-elements from effects due to stellar parameters, such as $T_{\rm eff}$, $\log g$, [Fe/H], 
$\xi$ \citep[see][]{Franchini04}; \cite{Jones95} found that the $H{\gamma}_{\rm HR}$--Fe4668 
diagram can be used to break the age-metallicity degeneracy in old stellar populations. Unfortunately, 
the H$_{\gamma}$ spectral line can be affected by emission producing unsatisfactory results 
\citep[see][]{Gibson99}.

In this Letter, we explore a high-resolution ({\it R}= 6000) simple stellar population (SSP) theoretical 
spectral energy distribution (SED) library searching for spectral indices useful for estimating the age 
and metallicity of stellar populations, and determine the signal-to-noise ratio (S/N) required of the 
observed spectrum of an intermediate-old ($> 100$ Myr) stellar population for our method to work. 
The structure of this Letter is as follows. In \S \ref{sec:models} we describe the library of SSP models. 
In \S \ref{sec:index-simulation} we present new spectral indices and establish their reliability to infer 
the age and metallicity of stellar populations. In \S \ref{sec:discussion} we discuss the results of our 
work.

\section{Theoretical models}
\label{sec:models}
In a separate paper (P. R. T. Coelho et al. 2020, in preparation, hereafter CBC20) we present a library 
of high-resolution SEDs of SSPs based on the PARSEC stellar evolutionary tracks \citep{Bressan12, Chen15} 
and the \cite{Coelho14} synthetic stellar spectral library \citep[see also][]{Coelho20}. The \cite{Coelho14} library covers the atmospheric parameter space from $\log T_{\rm eff}$=~3.5~to~4.3,  $\log g= -0.5$ to 
5.5, and 0.0017 $\le {\it Z} \le$ 0.049. For our purposes, three important characteristics of this library 
are 
{\it (a)} its high spectral resolution ({\it R}= 20,000), 
{\it (b)} its wavelength coverage from 2500 to 9000 {\AA}, and
{\it (c)} its wavelength sampling with step $\Delta\lambda = 0.02$ {\AA}.
In the CBC20 models, the time scale spans from 1 Myr to 15 Gyr, with a varying time step. All of these  
characteristics render the CBC20 models optimal for the development of tools necessary for the 
analysis of data obtained with high-resolution spectrometers like X-shooter \citep{Vernet11}, MEGARA 
\citep{Carrasco18} and MOSAIC \citep{Jagourel18}. For our analysis, we select three CBC20 models: 
a metal poor $\alpha$-enhanced ([$\alpha$/Fe]= 0.4) model, with constant iron abundance [Fe/H]= $-1.0$ 
(enhancement weighted metallicity {\it Z}= 0.0035, m10p04 in \citealt{Coelho14}'s notation), and two 
scaled solar models: one for solar metallicity ([Fe/H]= 0.0, {\it Z}= 0.017, p00p00) and a metal rich one 
([Fe/H]= 0.2, {\it Z}= 0.026, p02p00).

\begin{deluxetable*}{cccccc}
\tablecaption{\label{Table:indices}New indices defined following the Lick/IDS index definition (see text).}
\tablehead{
\mlc{1}{c}{Line}&\mlc{1}{c}{Index}& \mlc{1}{c}{Blue Band}        & \mlr{1}{*}{Index}& \mlc{1}{c}{Red Band}         &\mlc{1}{c}{Type}  \\
                &\mlc{1}{c}{ID}   & \mlc{1}{c}{\AA}              & \mlc{1}{c}{\AA}  & \mlc{1}{c}{\AA}&
}
\startdata
Ca 3933.66 & Ca3934  & 3906.0 -- 3912.0  & 3926.6 -- 3940.6  &  3948.0 -- 3954.0  & EW (\AA)  \\
Fe 4045.81 & Fe4045  & 4036.0 -- 4041.0  & 4044.8 -- 4046.8  &  4052.0 -- 4057.0  & EW (\AA)  \\
Mg 4481.13 & Mg4480  & 4473.0 -- 4478.0  & 4480.8 -- 4481.9  &  4484.0 -- 4489.0  & EW (\AA)   \\
\enddata
\end{deluxetable*}

\begin{figure}
\begin{center}
\includegraphics[scale=0.90]{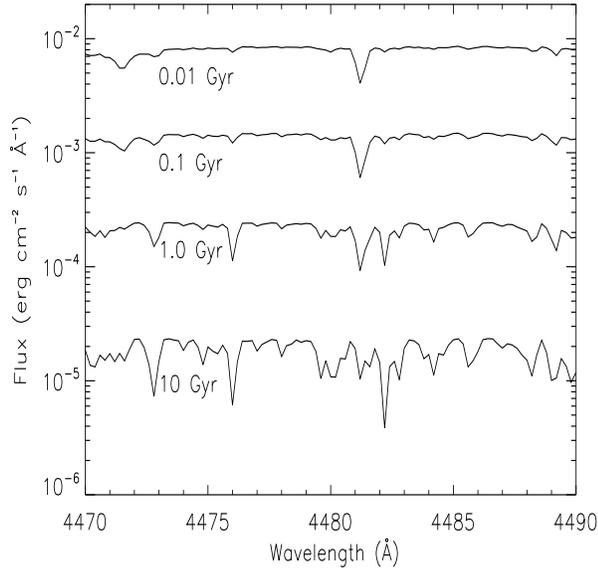}
\caption{Change of the Mg 4481.13 {\AA} line with age in our solar metallicity model. 
}
\label{fig:line_var}
\end{center}
\end{figure}

\begin{figure}
\begin{center}
\includegraphics[scale=0.85]{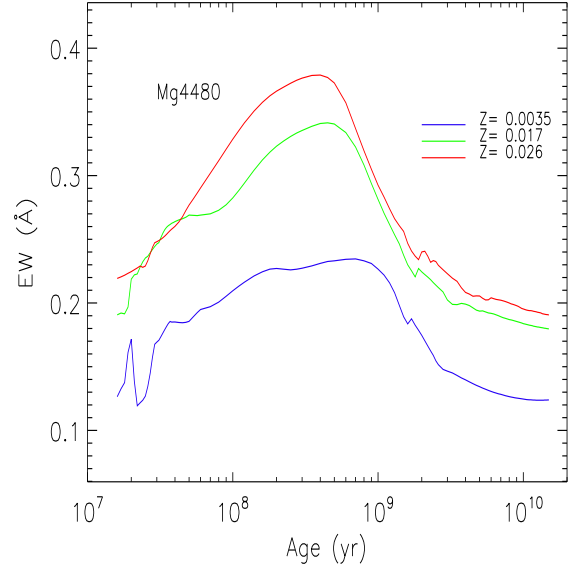}
\caption{Behavior of Mg4480 with age and metallicity.}
\label{fig:Indices_time}
\end{center}
\end{figure}

\section{Spectral line indices} 
\label{sec:index-simulation}

\begin{figure}
\begin{center}
\includegraphics[scale=0.95]{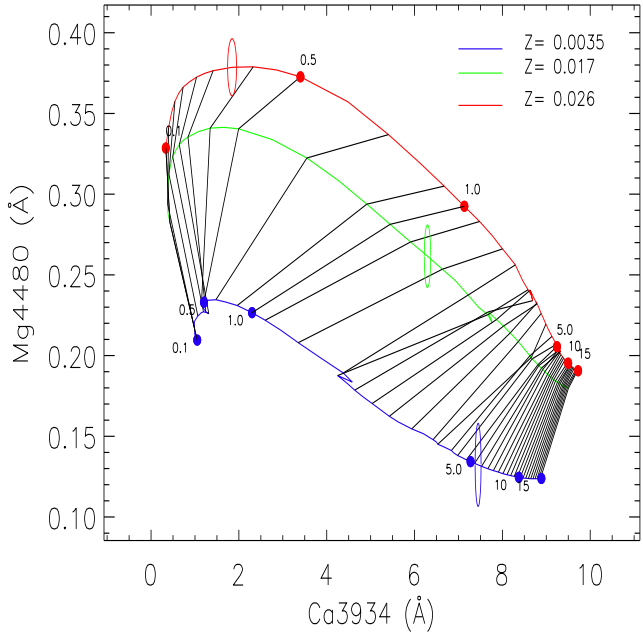}
\caption{Ca3934--Mg4480 index--index diagram. Red, green, and blue lines correspond to metal rich, solar 
metallicity and metal poor models, respectively. Numbers close to the dots indicate the age of the models 
in Gyr. The ellipses indicate the distributions of values estimated from the simulated observations for 
S/N = 10 (see the text).}
\label{fig:Index_Index_ellipse_Ca}
\end{center} 
\end{figure}

\begin{figure}
\begin{center}
\includegraphics[scale=0.95]{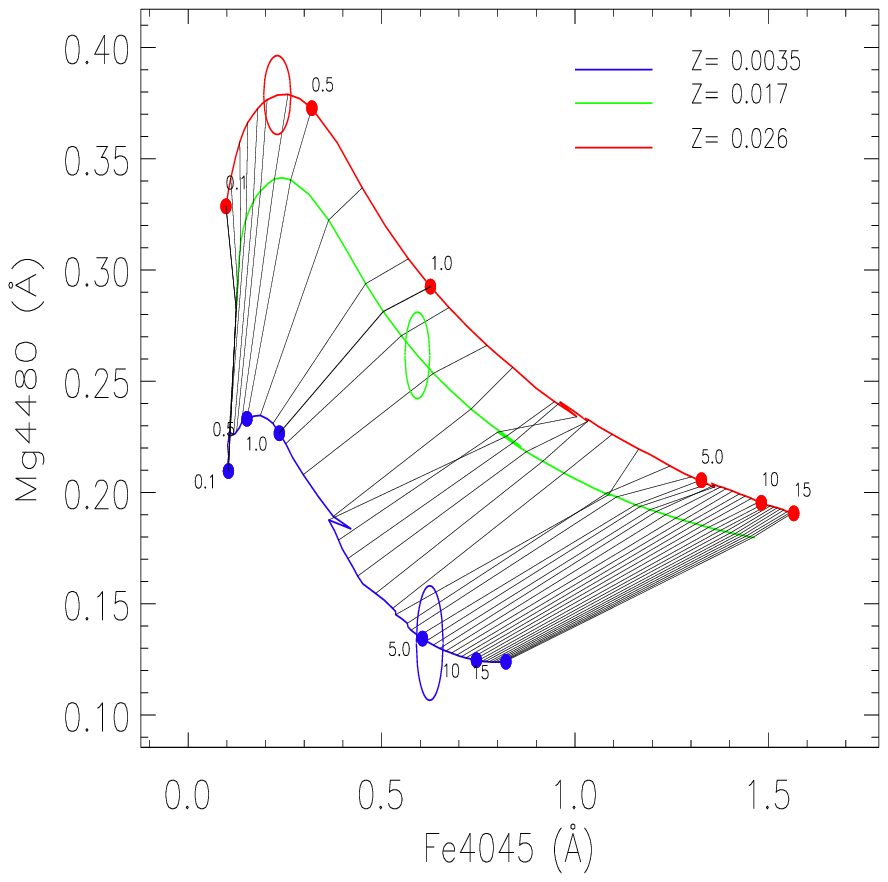}
\caption{Same as Figure \ref{fig:Index_Index_ellipse_Ca} but for the Fe4045--Mg4480 index--index diagram.}
\label{fig:Index_Index_ellipse_Mg}
\end{center} 
\end{figure}

We carefully inspected the SSP models, looking for strong atomic lines present in the spectra of stellar 
populations older than 100 Myr that are clearly distinguishable in the blue part of the spectrum 
(3655--5200 \AA). We chose this wavelength range because it is more populated by atomic lines and it is 
easily accessible to instruments mounted on ground-based telescopes. To carry out our analysis, we 
degraded and rebinned the theoretical SEDs to approximate instrumental properties of last-generation 
astronomical devices, e.g., {\it R}= 6000, $\Delta\lambda=\,0.2$ \AA. We identified several spectral lines 
that can be useful to study stellar populations. In this work, we will concentrate on the Ca 3933.66, 
Fe 4045.81 and the Mg 4481.13 {\AA} lines. We follow the standard Lick index rules \citep{Trager98} to 
define three new indices, Ca3934, Fe4045, and Mg4480. Each index involves a spectral line that is measured 
by its equivalent width
\begin{equation}
 EW=\int^{\lambda_2}_{\lambda_1}\left(1-\frac{F_{I\lambda}}{F_{C\lambda}}\right) d\lambda,
\end{equation}
where $\lambda_1$ and $\lambda_2$ are the limits of the wavelength interval where the index is defined, 
and F$_{I\lambda}$ and F$_{C\lambda}$ are the spectral feature and the continuum fluxes, respectively. 
The continuum flux is calculated by linear interpolation of mean fluxes of the pseudo-continua $F_p$ in
bands of around 5 {\AA} width in the blue and red sides of the main band. The pseudo-continuum is given by
\begin{equation}
 F_p=\frac{1}{{\lambda_2} - {\lambda_1}}\int^{\lambda_2}_{\lambda_1} F_{\lambda} d{\lambda},
\end{equation}
where $\lambda_1$ and $\lambda_2$ limit the wavelength interval defining the pseudo-continuum. In general, 
$\lambda_1$ and $\lambda_2$ are located close to the index, where there are no strong spectral lines. 
Table~\ref{Table:indices} lists the name of the index, the limits of the bands used to measure the index, 
and the pseudo-continua and the index type.

\begin{deluxetable*}{cccccccccc}
\tablecaption{\label{Table:fits}Fit results.}
\tablehead{
\mlc{1}{c}{Index--Index}  & \mlc{3}{c}{Mock Population}                        &  & \mlc{2}{c}{S/N = 100}           &  & \mlc{2}{c}{S/N = 10} \\
\mlc{1}{c}{Diagram}      & \mlc{1}{c}{ID} &  \mlc{1}{c}{Age (Gyr)} & \mlc{1}{c}{\it Z}  &  & \mlc{1}{c}{Age (Gyr)} & \mlc{1}{c}{\it Z\%} &  & \mlc{1}{c}{Age (Gyr)} & \mlc{1}{c}{\it Z\%}
}
\startdata
                & o-mp & 5.50 & 0.0035 & & $5.50\pm0.08$  & 100 & & $5.58\pm0.19$ &  100   \\
 Ca3934--Mg4480 & i-sm & 1.20 & 0.017  & & $1.20\pm0.02$  & 100 & & $1.20\pm0.27$ &  89.3  \\
                & y-mr & 0.35 & 0.026  & & $0.35\pm0.02$  & 100 & & $0.41\pm0.07$ &  54.6  \\
\hline
                & o-mp & 5.50 & 0.0035 & & $5.48\pm0.10$  & 100 & & $5.45\pm0.97$ &  100   \\
 Fe4045--Mg4480 & i-sm & 1.20 & 0.017  & & $1.20\pm0.02$  & 100 & & $1.16\pm0.08$ &  87.2  \\
                & y-mr & 0.35 & 0.026  & & $0.35\pm0.02$  & 100 & & $0.35\pm0.07$ &  77.2  \\
\enddata
\end{deluxetable*}

Figures \ref{fig:line_var} and \ref{fig:Indices_time} show, respectively, the behavior in time of the 
Mg 4481.13 {\AA} line in our solar metallicity model and the Mg4480 index\footnote{The strength 
of this index shows a peculiar behavior, increasing first and then decreasing. Instead, for the other 
two indices in Table~\ref{Table:indices} the strength of the index increases monotonically with time. 
Strictly speaking the CBC20 models are more reliable for age $\ge 30$ Myr, due to the coverage of the 
stellar library. Below this age there are some simplifications involved, and features at early ages as 
seen in Figure \ref{fig:Indices_time} maybe an artifact.} for different metallicity models. An index by 
itself is not useful to infer the age and the metallicity of an observed stellar population, since in 
all cases the indices suffer from the age-metallicity degeneracy; i.e., the value of the index for a 
young metal rich population can equal that for old metal poor population. Two index--index diagrams can 
be built with our three indices that prove useful for estimating age and metallicity. Figures 
\ref{fig:Index_Index_ellipse_Ca} and \ref{fig:Index_Index_ellipse_Mg} show the {\rm Ca3934--Mg4480} and 
the Fe4045--Mg4480 index--index diagrams, respectively. In these diagrams the two-dimensional grids 
indicate the change of the indices with age and metallicity. Points on the grid will have a unique 
associated age and metallicity. 

\begin{figure}
\begin{center}
\includegraphics[scale=0.8]{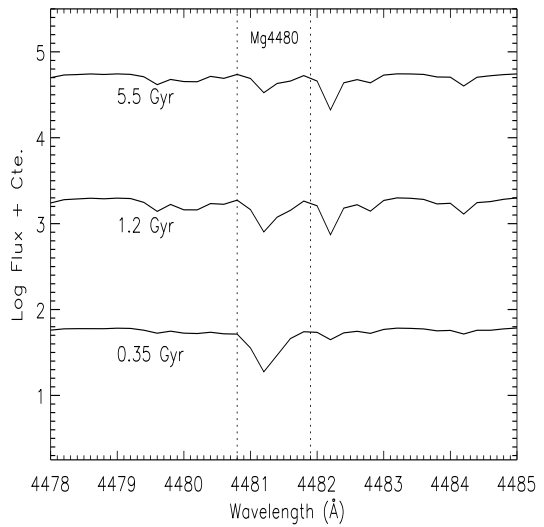}
\caption{Behavior of the Mg4480 line (inside dotted lines) with age and metallicity. SEDs for 
{\it Z}= (0.0035, 0.017, 0.026) at age= (5.5, 1.2, 0.35) Gyr, respectively, are shown.}
\label{fig:SED_Indices}
\end{center}
\end{figure}

To test the effectiveness of the index--index diagrams in determining the age and metallicity of stellar 
populations, we perform the following experiment. We select three spectra from the CBC20 models for a 
physically motivated combination of age and metallicity: 
(a) an old metal poor population of age 5.5 Gyr from the {\it Z}= 0.0035, [$\alpha$/Fe]= 0.4 model 
(hereafter o-mp),
(b) an intermediate age solar metallicity population of age 1.2 Gyr from the {\it Z}= 0.017 model 
(hereafter i-sm), and
(c) a young metal rich population of age 0.35 Gyr from the {\it Z}= 0.026 model (hereafter y-mr).
Figure \ref{fig:SED_Indices} shows the behavior of the selected spectra near the Mg4480 line.

We mimic observed spectra by adding Gaussian random noise to the flux with standard deviation 
$\sigma(\lambda)$ such that the resulting S/N = 10, 13.3, 20, 40, and 100 at all wavelengths, performing 
1000 realizations per S/N. Then we measure the spectral indices in the mock observations. Each pair of 
indices leads to a problem point in the index--index diagrams of Figures \ref{fig:Index_Index_ellipse_Ca} 
and \ref{fig:Index_Index_ellipse_Mg}. To this point we associate the (age, metallicity) of the model at 
the closest corner in the corresponding grid. In general, if the problem point falls in a region where 
the mesh is well separated, as it happens for the constant {\it Z} lines, we recover the true value. 
The distributions of our estimates for S/N = 10 are indicated in Figures \ref{fig:Index_Index_ellipse_Ca} 
and \ref{fig:Index_Index_ellipse_Mg} by ellipses. The mean values define the centers of the ellipses and 
the standard deviations their axes. Table~\ref{Table:fits} lists our results for S/N= 100 and 10. 
The values in the columns marked {\it Z}\% represent the percentage of the time that we recover the true 
{\it Z}. In the S/N = 100 case, our estimates are very close to the true values. For S/N = 10 our estimates 
agree with the true values within errors. We conclude that the (age, {\it Z}) estimates obtained from both 
index--index diagrams are quite close to the true values and that these diagrams provide a promising tool 
to determine these parameters.

\begin{figure}
\begin{center}
\includegraphics[scale=0.95]{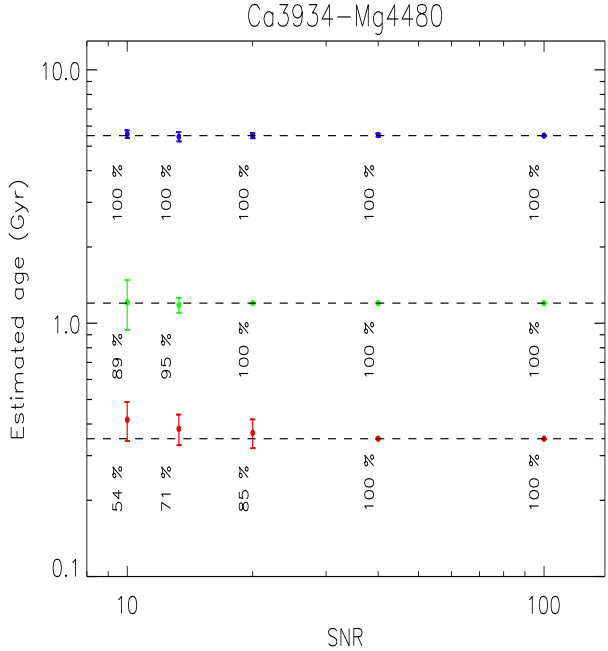}
\caption{Age estimates from the Ca3934--Mg4480 index--index diagram vs. the S/N of the mock observations 
for the o-mp (blue), i-sm (green), and y-mr (red) populations. Dashed lines represent the true age. The 
percentage of results providing the true {\it Z} are indicated.}
\label{fig:Resultados_Edad_Ca}
\end{center} 
\end{figure}

\begin{figure}
\begin{center}
\includegraphics[scale=0.95]{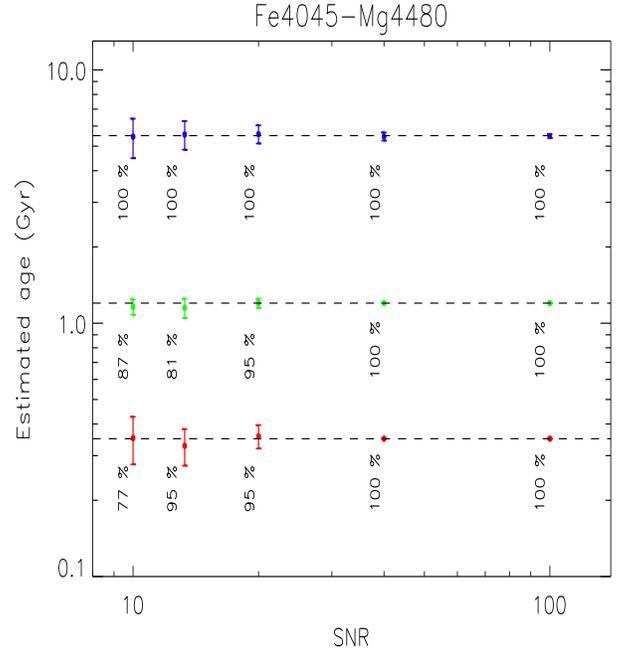}
\caption{Same as Figure \ref{fig:Resultados_Edad_Ca} but for the Fe4045--Mg4480 index--index diagram.}
\label{fig:Resultados_Edad_Mg}
\end{center} 
\end{figure}

Figures \ref{fig:Resultados_Edad_Ca} and \ref{fig:Resultados_Edad_Mg} show the behavior of our age 
estimates from the Ca3934--Mg4480 and Fe4045--Mg4480 index--index diagrams with S/N, respectively.
For large SNR, the uncertainty in the estimated age is very small, but it increases considerably for 
low S/N.

\begin{figure}
\begin{center}
\includegraphics[scale=0.95]{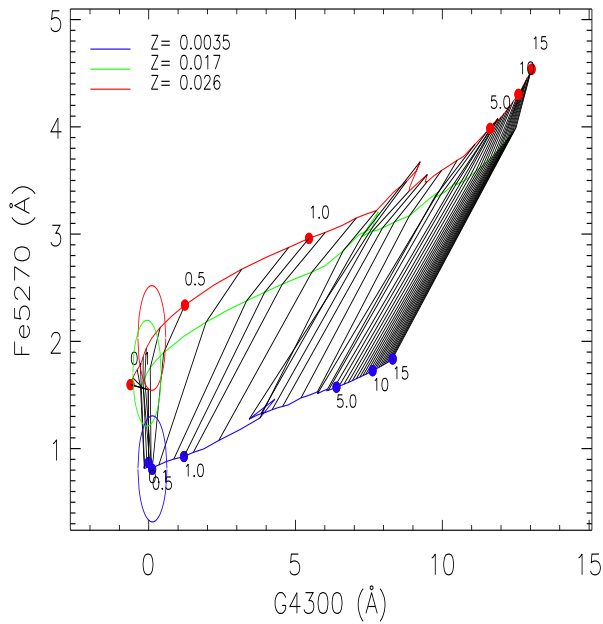}
\caption{Same as Figure \ref{fig:Index_Index_ellipse_Ca} but for the Lick G4300--Fe5270 index--index 
diagram obtained from degraded and rebinned model SEDs (see the text).}
\label{fig:index_index_Lick}
\end{center} 
\end{figure}

For comparison, we repeat our tests using the full set of 21 Lick spectral indices. The model SEDs were 
degraded to a resolution with FWHM = 3 {\AA} and resampled to $\Delta\lambda = 1.0$ {\AA}.We added 
Gaussian random noise to the 0.35~Gyr SED to mimic an observed spectrum. We measured the indices using 
the most up-to-date wavelength definitions\footnote{http://astro.wsu.edu/worthey/html/index.table.html} 
and found a few index--index diagrams that might be used to estimate the basic parameters of intermediate- 
to old-age stellar populations. Figure \ref{fig:index_index_Lick} displays the G4300--Fe5270 index--index 
diagram. The ellipses correspond to the distributions of our estimates for the S/N = 10 case. The 
uncertainty in the estimated age grows so quickly with decreasing S/N that the true values are recovered 
only for the highest S/N (100).

\section{Discussion}
\label{sec:discussion}
We analyze a library of theoretical SEDs calculated at high resolution with the goal of identifying 
spectral features that can be used to determine the basic parameters of intermediate to old stellar 
populations. We define three new spectral indices in the blue part of the spectrum, which proved useful 
for simultaneously estimating age and {\it Z} for stellar populations older than 100 Myr. To test the 
efficiency of this tool we simulate thousands of spectroscopic observations by adding Gaussian random 
noise to model SEDs of intermediate to old age and different {\it Z}. The new spectral indices measured 
on each mock spectrum were used in conjunction with the new index--index diagrams to estimate age and 
{\it Z} for the mock population. We find that for S/N $\ge 10$ we recover age and {\it Z} with relatively 
high accuracy. The new index--index diagrams allow to break the age--metallicity degeneracy present in 
stellar populations, as long as the indices are measured in high-resolution SEDs ({\it R}= 6000) of 
S/N $\ge 10$.

A similar analysis was performed employing the full set of 21 Lick indices. We find that in this case 
the uncertainty in the recovered age and Z grows so quickly with decreasing S/N that parameter estimates 
are reliable only for the highest S/N (100 in our tests).

It is important to keep in mind that each pair of spectral indices displays a particular distribution 
in an index--index diagram. Therefore, some indices are useful to estimate parameters of intermediate-age 
populations, while other indices are more suitable for old populations. Here we report results for 
a limited number of indices and index--index diagrams; having both Ca, Mg and Fe indices we show 
that the age and metallicity estimated do not strongly depend on the $\alpha$-enhanced elements 
abundances. In a forthcoming paper we will describe a larger set of spectral indices and index--index 
diagrams in the 3655--5200~\AA\ wavelength range. These indices can be used to estimate the basic 
parameters of intermediate to old stellar populations.

\section{Acknowledgements}
We thank the anonymous referee for a quick review of this manuscript. 
Y.D.M. thanks CONACyT for the research grant CB-A1-S-25070. 
G.B. acknowledges financial support from the National Autonomous University of Mexico (UNAM) through grant 
DGAPA/PAPIIT IG100319 and from CONACyT through grant CB2015-252364. 
P.C. acknowledges support from Conselho Nacional de Desenvolvimento Cient\'ifico e Tecnol\'ogico
(CNPq 310041/2018-0). 
P.C. and G.B. acknowledge support from Funda\c c\~ao de Amparo \`a Pesquisa do Estado de S\~ao Paulo 
through projects FAPESP 2017/02375-2 and 2018/05392-8. P.C. and S.C. acknowledge support from USP-COFECUB 
2018.1.241.1.8-40449YB.
A.G.d.P. acknowledges financial support from the Spanish MCIUN under grant no. RTI2018-096188-B-I00

\end{document}